\documentclass{ws-procs9x6}

\begin{document}

\begin{center}
{\bf\large $\eta$ photoproduction off the neutron at GRAAL:} \\
{\it\large Evidence for a resonant structure at W = 1.67~GeV}  

\vspace*{0.5cm}

V.Kuznetsov$^1$\footnote{E-mail \uppercase{S}lava@cpc.inr.ac.ru,
\uppercase{S}lavaK@jlab.org},
O. Bartalini$^2$, V. Bellini$^3$, M. Castoldi$^4$, A. D'Angelo$^2$,
J-P. Didelez$^5$, R. Di Salvo$^2$, A. Fantini$^2$, D. Franco$^2$, G. Gervino$^6$,
F. Ghio$^7$,
B. Girolami$^7$, A. Giusa$^3$, M. Guidal$^5$, E. Hourany$^5$, R. Kunne$^5$,
A. Lapik$^1$, P. Levi Sandri$^8$, D. Moricciani$^2$,
L. Nicoletti$^3$, C. Randieri$^3$, N. Rudnev$^{9}$, G.
Russo$^3$, C. Schaerf$^2$, M.-L. Sperduto$^3$, M.-C. Sutera$^3$, A.
Turinge$^{10}$.

\vspace{0.5cm}

{\it $^1$Institute for Nuclear Research, 117312 Moscow, Russia}\\
{\it $^2$INFN sezione di Roma II and Universit\`a di Roma "Tor Vergata",
00133 Roma, Italy}\\
{\it $^3$INFN Laboratori Nazionali del Sud and Universit\`a di Catania,
95123 Catania,Italy}\\
{\it $^4$INFN Genova and Universit\`a di Genova, 16146 Genova, Italy }\\
{\it $^5$IN2P3, Institut de Physique Nucl\'eaire, 91406 Orsay, France}\\
{\it $^6$INFN sezione di Torino  and Universit\`a di Torino, 10125 Torino,Italy}\\
{\it $^7$INFN sezione Sanit\`a and Istituto Superiore di Sanit\`a,00161 Roma, Italy}\\
{\it $^8$INFN Laboratori Nazionali di Frascati, 00044 Frascati, Italy}\\
{\it$^{9}$Institute of Theoretical and Experimental Physics, Moscow, Russia}\\
{\it $^{10}$RRC "Kurchatov Institute", Moscow, Russia}

\end{center}

\abstracts{ New (preliminary) data on $\eta $ photoproduction 
on the neutron are presented. These data reveal a resonant 
structure at W=1.67 GeV. }

Meson photoproduction on the neutron may provide essentially new 
information regarding the spectrum of baryons. An example is given 
by a model\cite{mok} which exploits the SU(6) symmetry and assumes 
single-quark transitions from ground nucleons to the $[70,1^-]$ 
supermultiplet. The model predicts only weak photoexcitation of the 
$D_{15}(1675)$ resonance from the proton target. Conversely, 
photon-neutron couplings of $D_{15}(1675)$ calculated in the 
framework of this approach are not small. Measurements of the 
relative strength of photoneutron/photoproton interaction are 
therefore an important testing ground for this (and others) 
theoretical approaches.

Another example is possible photoexcitation of the non-strange
pentaquark state, which is associated with the second member of 
an antidecuplet of exotic baryons\cite{diak,jafw}. Evidence for the 
lightest member of the antidecuplet, the $\Theta^+(1540)$ baryon, 
is now being widely discussed\cite{nspr}. It can be produced, in
particular, by photoexcitation of the nucleon. However, exact 
$SU(3)_F$ would forbid the proton photoexcitation into the 
proton-like antidecuplet member. The chiral soliton model predicts 
that photoexcitation of the non-strange pentaquark has to be 
suppressed on the proton  and should occur mainly on the neutron, 
even after accounting for $SU(3)_F$ violation\cite{max}. Estimates 
of the mass and width of the non-strange pentaquark are ambiguous. 
As initial input, the mass was used to be 1.71~GeV\cite{diak}, with 
the width estimated $\sim 40$~MeV. More recent evaluation of the 
chiral soliton approach led to the range of 1.65 -- 
1.69~GeV\cite{diak1}.  In the di-quarks approach\cite{jafw}, the 
mass of the pentaquark with hidden strangeness is quoted about 
1.7~GeV.
Modified partial wave analysis of the 
$\pi N$ scattering\cite{str} suggests two possible candidates, at 
1.68~GeV and/or at 1.73~GeV, with the total width about 10~MeV and 
the partial width for $\pi N$ decay mode less than 0.5~MeV. Thus, 
photo-neutron excitation data, and their comparison with photo-proton 
excitation, may be important both in establishing existence of 
pentaquarks and in discriminating between different theoretical 
concepts.  
Among other reactions, $\eta$ photoproduction has been suggested  
as particularly sensitive to the manifestation of the non-strange 
pentaquark\cite{diak,jafw,max,str}.

Up to now, $\eta$ photoproduction on the neutron has been explored 
only in the region of the $S_{11}(1535)$ resonance from threshold
up to W = 1.6~GeV\cite{inc,exc,bonn}.  Some of the previous 
experiments\cite{inc} were limited to inclusive measurements 
detecting only the outgoing $\eta$.  In exclusive 
experiments\cite{exc,bonn}, both the $\eta$ and the recoil nucleon 
are detected. This makes it possible to discriminate between $\eta n$ 
and $\eta p$ final states and to select events corresponding to 
quasi-free kinematics. 
 
A new exclusive measurement has been performed at 
GRAAL\cite{gra1,gra2}
using a deuteron target. Both quasi-free $\gamma n \rightarrow \eta 
n$ and $\gamma p \rightarrow \eta p$ reactions were explored 
simultaneously in the same experimental run under the same conditions 
and solid angle.  Two photons from $\eta \rightarrow 2\gamma$ decay 
were  detected in the BGO crystal ball\cite{bgo}.  Recoil neutrons 
and protons emitted at $\Theta_{lab} = 3 - 23^{\circ}$, were detected 
in an assembly of forward detectors, which includes two planar 
multiwire chambers, a time-of-flight (TOF) wall made of thin 
scintillator strips, and a lead-scintillator sandwich TOF 
wall\cite{rw}. 
The latter detector adds the option of neutron detection with an 
efficiency of $\sim 22\%$. The momenta of the $\eta$ and recoil 
nucleons were reconstructed from measured energies, TOFs and angles 
of outgoing particles.  

\begin{figure}[ht]
\centerline{\epsfxsize=2.5in\epsfbox{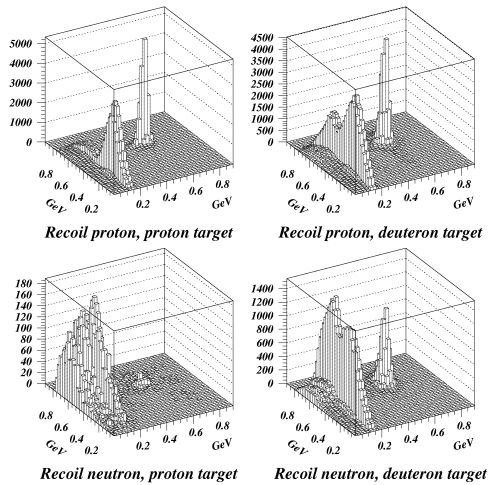}}   
\caption{ Bi-dimensional plots of invariant mass of two photons versus
missing mass calculated from momenta of recoil nucleons 
for proton and deuteron targets. \label{Fig1}}
\end{figure}
 
As a first step, the identification of the $\eta n$ and $\eta p$ 
final states was achieved in a way similar to that used in the 
previous measurements\cite{gra2} on the free proton. The $\eta$ was 
identified by means of the invariant mass of two photons and its 
momentum was reconstructed from measured photon energies and angles. 
Then measured parameters of the recoil nucleon were compared with 
ones expected assuming quasi-free kinematics.  Fig.~1 shows 
bi-dimensional plots of $2\gamma$ invariant mass versus $\eta$ 
missing mass obtained in experimental runs with proton and deuteron 
targets.  A good $\eta p$ signal was obtained with the proton target, 
while only  few $\eta n$ events were detected in this run. Signals of 
both final states clearly appear with the deuteron target.

In case of a photon interaction with the nucleon bound in a deuteron 
target, event kinematics is ``peaked" around that one on the free 
nucleon.
Fermi motion of the target nucleon changes the effective energy of 
photon-nucleon interaction and affects parameters of outgoing 
particles. Part of events may suffer from re-scattering and 
final-state interaction\cite{kudr}.  The goal of the second-level 
selection was to reduce re-scattering events, the remaining 
background (mostly from $\gamma d\to\eta\pi NN$), and those events  
whose kinematics are strongly distorted by Fermi motion. Additional 
cuts on the recoil nucleon missing mass $M(\gamma N, \eta)$ and 
$\Delta$TOF have been applied. Those events in which the detection 
of two photons and the recoil nucleon was accompanied by the detection 
of any low-energy particle(s) in the $4\pi$ GRAAL detector, have been 
eliminated from the analysis. 

The strategy at this stage was to study the dependence of spectra
of selected events on cuts. Two criteria of quality of the selection 
procedure have been exploited: 
(i) distribution of Fermi momentum of the target neutron
reconstructed as ``missing momentum" with small correction 
on binding energy; 
(ii) difference of the center-of-mass energy W calculated from
the momentum of the initial-state photon and assuming the target 
nucleon at rest, and the center-of-mass energy deduced as the 
invariant mass of the final-state $\eta$ and the neutron.  The first 
quantity includes uncertainties due to Fermi motion and is ``peaked" 
around  the real center-of-mass energy of photon-nucleon interaction. 
The $\eta n$ invariant mass is not affected by Fermi motion but 
includes large uncertanties (50 -- 80~MeV, FWHM) due to detector 
resolution.

\begin{figure}[ht]
\centerline{\epsfxsize=4.5in\epsfbox{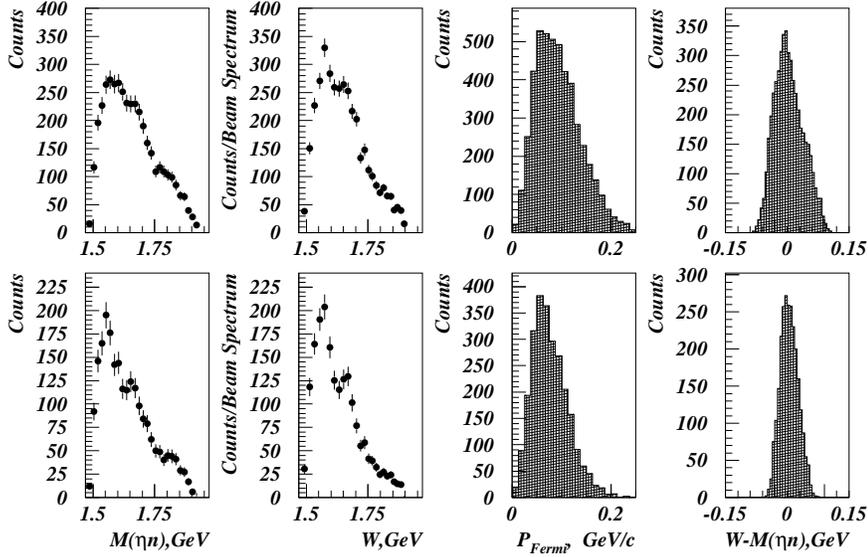}}   
\caption{Spectra of center-of-mass energy, calculated as invariant 
mass of
final-state $\eta$ and the neutron (first column), from the energy of 
the 
itial-state photon and assumimng the target neutron in the rest 
(second column), 
Fermi momentum of the target neutron (third
column), and difference between initial-state and final-state 
center-of-mass energies (fourth column)
after different cuts (see text). \protect\label{Fig2}}
\end{figure}
 
In the upper row of Fig.~2, the final-state (first column) 
and initial-state (second column) W spectra obtained with 
the first-level cut are shown. Both of them indicate a wide bump in 
the region 1.6 -- 1.7~GeV. The Fermi momentum (third column) exhibits 
a broad distribution.  Plots in the lower row correspond to final 
cuts.  Both final and initial-state spectra are similar and show an 
enhancement of the $S_{11}(1535)$ resonance below 1.6~GeV. The bump 
near 1.67~GeV observed in the previous spectra, becomes more narrow 
and well-pronounced. The Fermi-momentum spectrum is more compressed 
and has its maximum near 0.05~GeV/c, as expected for quasi-free 
events. 

Evolution of spectra in Fig.~2 suggests that most of events rejected 
by the second-level cuts 
either originate from re-scattering and final-state interaction or 
strongly suffer from Fermi motion. On the contrary, events shown in 
lower-row plots, are more ``clean".  They correspond to quasi-free 
photoproduction, and the distortion due to Fermi motion is reduced. 
The latter fact makes it possible to clearly reveal the structure at 
1.67~GeV.
These events were found suitable for the further analysis.

\begin{figure}[ht]
\vspace*{-0.2 cm}
\centerline{\epsfxsize=3.8in\epsfbox{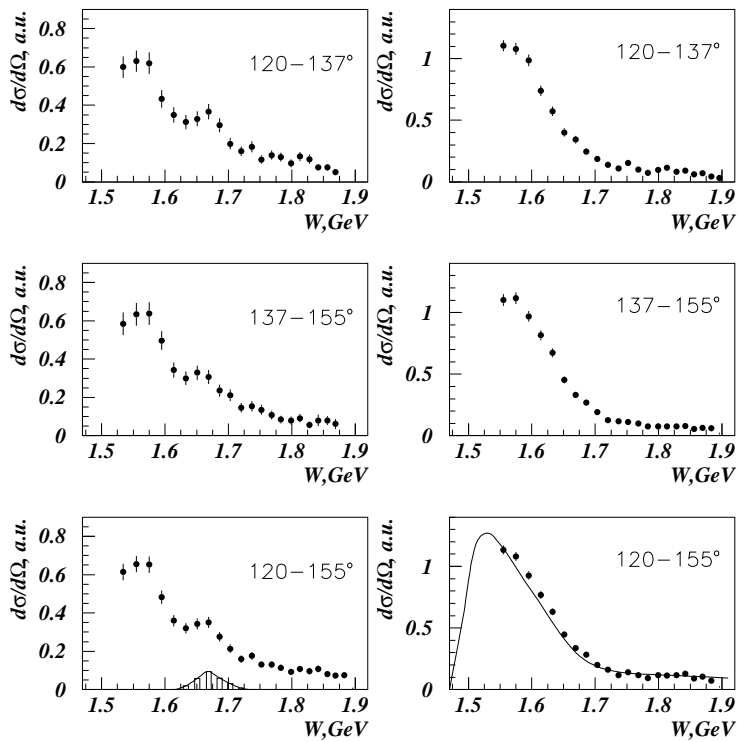}}   
\caption{Preliminary quasi-free $\eta n$(left) and $\eta p$(right) 
photoproduction cross sections (dark circles). Solid line indicates 
E429
solition of the SAID $\gamma p\to\eta p$ partial wave 
analysis\protect\cite{str2}. Dashed area shows simulated contribution 
of a narrow state at W=1.675 GeV.}
\end{figure}

Preliminary quasi-free $\eta n$ and $\eta p$ photoproduction cross 
sections 
are shown in Fig.~3. The normalization has been done by comparing 
quasi-free proton data and the E429 solution of the SAID $\gamma 
p\to\eta p$ partial wave analysis\cite{str2} for $\eta$ 
photoproduction on the proton which was obtained from the fit to all 
available data including recent data from GRAAL\cite{gra1,gra2}, 
JLab\cite{pas}, and SAPHIR\cite{saph}.  Such normalization made it 
possible to avoid ambiguities related to the fact that some of the 
events 
are lost due to the re-scattering and final-state interaction.  Above 
1.55~GeV, a reasonable coincidence in the shape of the cross sections 
has been 
obtained. At lower energies, re-scattering and final-state interaction 
become more significant and play a dominant role near 
threshold\cite{kudr}.  Error bars shown in Fig.~3 correspond to  
statistical uncertainties only. The present normalization uncertainty 
of 12\% 
originates mostly from the quality of simulations of 
quasi-free processes and from uncertainties in the neutron detection 
efficiency. 
\begin{figure}[ht]
\vspace*{-0.5 cm}
\centerline{\epsfxsize=3.8in\epsfbox{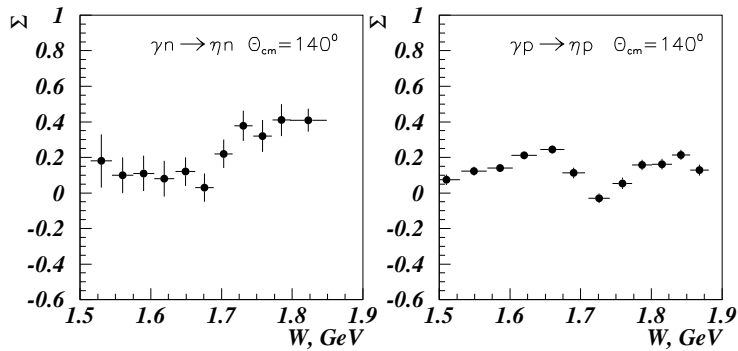}}   
\caption{Beam asymmetry $\Sigma$ for $\eta n$(left) and $\eta 
p$(right)
photoproduction. \protect\label{Fig4}}
\end{figure}

At ~W below 1.6~GeV, both cross sections exhibit bumps due to the 
$S_{11}(1535)$ resonance.  At higher ~W, an additional 
structure clearly appears for the neutron and is not seen on the 
proton. Remarkably, beam asymmetry $\Sigma$ (Fig.~4) shows 
pecularities at the same energies. In the region of $S_{11}(1535)$ 
resonance,  $\Sigma$ ranges around 0.2 and is nearly the same for the 
neutron and the proton. In the region 1.65 -- 1.73~GeV, there are 
step-like changes.  The trends of these changes are opposite: for the 
proton, the asymmetry becomes almost 0, while for the neutron it rises 
up to 0.4.  It is worth noting that beam asymmetry is more sensitive 
to the non-dominant contributions than cross section since it is 
given by the interference of helicity amplitudes $H_i$\cite{str1} 
corresponding to four possible helicity states of the target and 
recoil nucleon:
 
 \vspace{0.2 cm}

 \centerline{$d\sigma /d\Omega \sim \mid H_{1}\mid^{2} + \mid 
H_{2}\mid^{2} + 
 \mid H_{3}\mid^{2} + \mid H_{4}\mid^{2}$}

 \vspace{0.2 cm}

 \centerline{$\Sigma \sim Re(H_{1}H_{4}^{*}-H_{2}H_{3}^{*}) $}
 \vspace{0.2 cm}

Both the observed peak in the cross section on the neutron and 
corresponding changes in beam asymmetry might be an indication that 
one of the nucleon resonances has much stronger photocoupling to the 
neutron than to the proton.  In Fig.~3, the simulated contribution of 
a narrow (10~MeV) resonance state with a mass of 1.675~GeV is shown. 
Such a state appears in the cross section as a 40~MeV wide peak due to 
Fermi motion of the target nucleon.  The shape of the simulated peak 
fits quite well the shape of the peak observed in the cross section 
on the neutron.

Therefore, this peak may be a signal of a relatively narrow state. 
Potentially, this state looks promising as a candidate for 
the non-strange pentaquark. The feature of strong photocoupling to 
the neutron agrees with the prediction of the chiral soliton 
model\cite{max}. On the other hand, one cannot exclude that the 
observed peak is a manifestation of one of the known resonances, 
in particular, the $D_{15}(1675)$, as is suggested by the single-quark 
transition model\cite{mok}. A crucial task is to ``unfold" the 
cross-section and beam-asymmetry data from Fermi motion, in order 
to achieve a reliable estimate of the width of this state.  As a 
further step, a partial wave analysis will be needed to fix its 
quantum numbers.

It is worth to add that the kaon photoproduction has been  quoted as 
well to be particularly sensitive to the signal of the non-strange  
pentaquark\cite{diak,jafw,max,str} as well. Very preliminary 
indications on 
a state at 1.72~GeV have been obtained in $\gamma n \rightarrow 
K^0_s\Lambda$
and $\gamma n \rightarrow K^+\Sigma^-$ reactions\cite{grakl},
in production of $K^0_s\Lambda$ final state 
in $Au+Au$ collision\cite{kabana}, and in
$pp \rightarrow  K^+\Lambda p$ reaction\cite{cosykl}.

This work was supported by the Universit\`a di Catania and Laboratori
Nazionale del Zud, INFN Sezione di Catania (Italy).
Discussions with Ya.~Azimov, W.~Briscoe, V.~Burkert, D.~Diakonov, 
M.~Kotulla,
B.~Krusche, A.~Kudryavtsev, V.~Mokeev, E.~Pasyuk, M.~Polyakov, 
A.~Sibirtsev, 
I.~Strakovsky, and R.~Workman were very helpful.

\end{document}